\documentclass[manuscript]{aastex63}

\begin{document}

\newcommand{\kms}{~km$\,$s$^{-1}$}
\title{A JWST Preview: Adaptive-Optics Images of H$_2$, Br-$\gamma$, and K-continuum in Carina's Western Wall}

\correspondingauthor{Patrick Hartigan}
\email{hartigan@sparky.rice.edu}

\author[0000-0002-5380-549X]{Patrick Hartigan}
\affiliation{Physics and Astronomy Dept., Rice University, 6100 S. Main, Houston, TX 77005-1892}

\author[0000-0002-7639-5446]{Turlough Downes}
\affiliation{Centre for Astrophysics \& Relativity, School of Mathematical Sciences, Dublin City University, Glasnevin, Dublin 9, Ireland}

\author[0000-0001-8061-2207]{Andrea Isella}
\affiliation{Physics and Astronomy Dept., Rice University, 6100 S. Main, Houston, TX 77005-1892}

\begin{abstract}

We present the first wide-field near-infrared adaptive optics images of Carina's Western Wall (G287.38-0.62),
one of the brightest and most well-defined irradiated interfaces known in a region of massive star
formation.  The new narrowband H$_2$ 2.12$\mu$m, Br-$\gamma$ and
K-continuum images from Gemini South trace the photoevaporative flow from the cloud and identify locations
where UV-radiation from the surrounding massive stars excites molecular hydrogen to fluoresce.
With a field of view of $\sim$ 1.5\arcmin\ $\times$ 2.9\arcmin\ and spatial resolution between 60 $-$ 110 mas, the
new images show a spectacular level of detail over a large area, and presage what JWST should achieve.
The Wall is convex in shape, with a large triangular-shaped extension near
its apex. The interface near the apex consists of 3 $-$ 4
regularly-spaced ridges with projected spacings of $\sim$ 2000~AU, suggestive of a large-scale
dynamically-important magnetic field. The northern edge of the Wall breaks into several swept-back
fragments of width $\sim$ 1800~AU that resemble Kelvin-Helmholtz instabilities, and the
southern part of the Wall also shows complex morphologies including a sinusoidal-like variation with
a half-wavelength of 2500~AU.  Though the dissociation front must increase the density along the
surface of the Wall, it does not resolve into pillars that point back to the ionization sources,
as could occur if the front triggered new stars to form. 
We discovered that MHO 1630, an H$_2$ outflow with no clear driving source
in the northern portion of the Wall,
consists of a series of bow shocks arrayed in a line.

\end{abstract}


\section{Introduction} \label{sec:intro}

Star formation occurs both within low-mass dark clouds and inside large molecular
cloud complexes, but the role radiation plays in the two types of regions
differs markedly \citep[e.g.][for a review]{tan14}.  Unlike their more quiescent counterparts,
giant molecular clouds typically form O and B stars, and ultraviolet
radiation from these massive stars injects enough energy into the surrounding
medium to create large H~II regions. The resulting volumes of expanding ionized gas remove
material that might otherwise accrete onto a protostar, while radiation photoevaporates
the outer portions of disks and may alter the spectrum of
stellar masses created in the region \citep{winter18,krumholz11}.  On the other hand,
density waves driven into molecular clouds by radiation from massive stars
could also potentially trigger new stars to form, either directly as the fronts
compress existing clumps into gravitationally
unstable cores, or indirectly if the fronts increase the overall density in a cloud to the
point where subsequent collisions between cloud fragments induce collapse \citep[e.g.][]{haworth11}.
Radiative feedback and winds from newly-formed stars into the nascent cloud material also
drive structures on scales of tens or even hundreds of pc, and create
sequential episodes of star formation \citep[e.g.][]{venuti18} and clear out chimneys and bubbles
perpendicular to the galactic plane \citep{rob18}. On smaller scales, 
how massive stars form remains a topic of intense theoretical work, as
models must include the effects of strong radiation pressure and photoevaporation
from massive protostars that both limit the available timescale for accretion 
and may widen existing outflow cavities \citep{tanaka17,rosen16}.

To understand quantitatively how intense radiation fields from young massive stars affect star formation
in molecular clouds, we must assess parsec scales, where filaments respond to the
influence of large-scale magnetic fields, smaller scales of order 0.1~pc relevant to
core formation and fragmentation, and protostellar disk scales of several hundred AU \citep[e.g.][]{li18}.
The most intense radiation environments occur in regions that produce early O-stars, and
even the closest of these are over 1.5~kpc away, so to resolve structures on $\sim$ 100~AU
scales requires spatial resolutions better than 0.1\arcsec . This increase in resolution 
of an order of magnitude over typical non-adaptive-optics images holds the key to bridging 
interstellar scales to Oort cloud scales.

Recent advances in instrumentation, especially with adaptive optics (AO), have made it
possible to acquire nearly diffraction-limited images in the near-infrared over
a field of view greater than an arcminute in size. Such observations are ideal
for tracing large-scale outflows and shock
waves in star-forming regions where subarcsecond resolution uncovers morpologies
crucial to interpreting the physics at work in these systems
\citep[e.g.][]{bally15}. Wide-field near-infrared AO images presage what
the James Webb Space Telescope (JWST) should deliver when it becomes operational,
and complement the highest-resolution ALMA maps of molecular cloud filaments and cores,
which also have $\lesssim$ 1\arcsec\ resolution \citep[e.g.][]{cheng18}.
 
\begin{figure}
\plotone{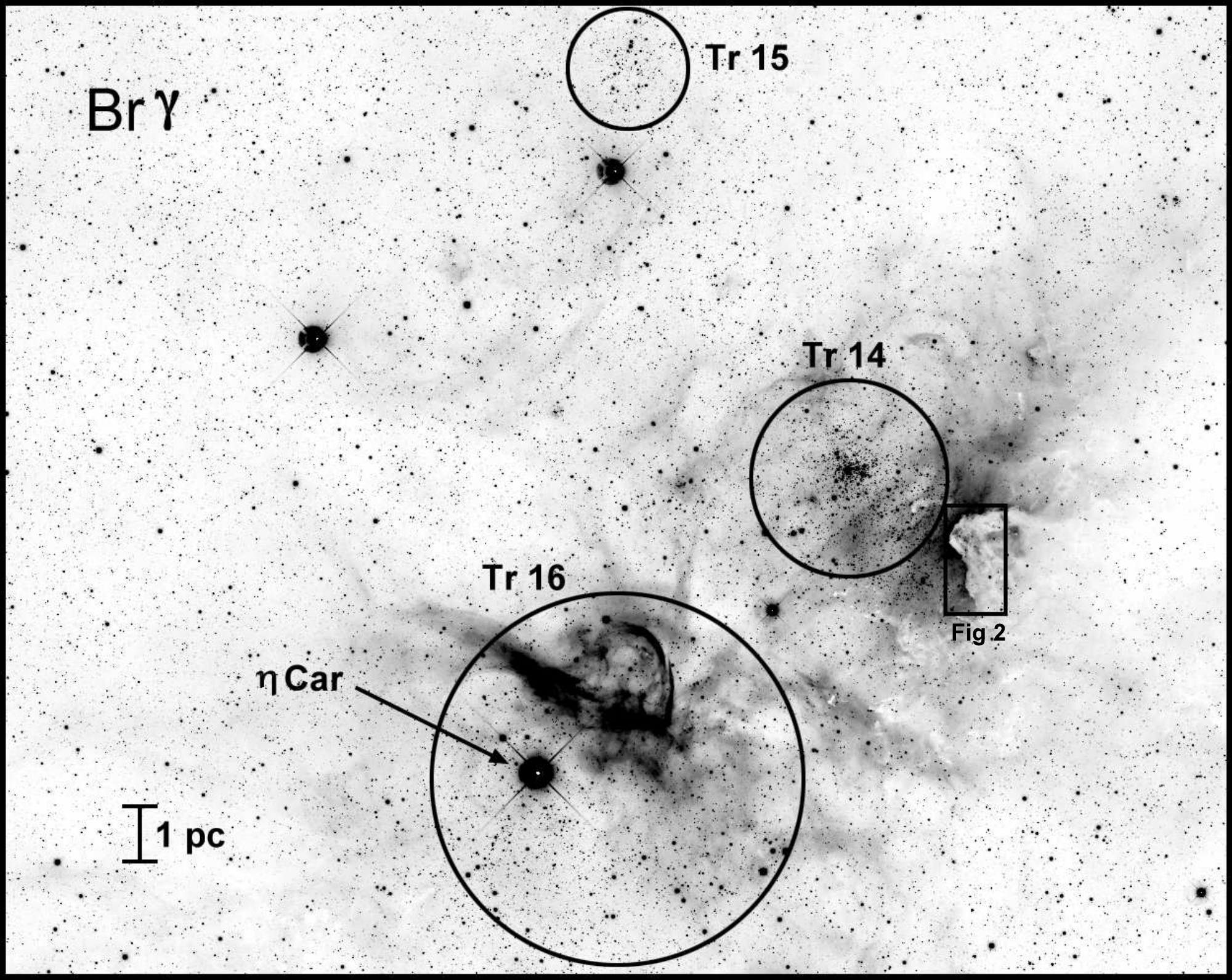}
\caption{Br-gamma image of the central portion of the Carina star formation region
from \citep{hartigan15}. Locations of the Trumpler 14, 15, and 16 clusters are circled. The boxed
area shows the region imaged in Fig.~\ref{fig:overview}.}
\label{fig:brg}
\end{figure}

In this paper we use the GSAOI adaptive optics
imager on Gemini to study the Western Wall (G287.38-0.62) in Carina,
one of the nearest \citep[2.3 kpc; ][]{lim19}
and most strongly-irradiated interfaces known to occur in a region of massive star
formation. This interface appears as an unassuming dark cloud at optical wavelengths,
but it stands out as the brightest feature in the entire Carina complex when imaged
in the 2.12 $\mu$m H$_2$ and Br-gamma emission lines \citep[][Fig.~\ref{fig:brg}]{hartigan15}.
The H$_2$ 2.12 $\mu$m line is excited by UV fluorescence, and provides
a means to trace irradiated interfaces in regions like Carina where UV emission lines
are absorbed by dust along the line of sight.  The observed spatial offsets
between H$_2$ and Br-$\gamma$ emission, the latter
arising in a photoevaporative flow as UV radiation
ionizes hydrogen, generally agree well with theoretical model predictions \citep{ch18}.
Over a dozen O-stars in the nearby open cluster Tr~14 and a similar number in the cluster Tr~16,
including an O2 star (HD 93129A) and an O3.5 star (HD~93129B) in Tr~14 and an O2.5 star (HD~93162)
and O3.5 star (ALS~15210) in Tr~16, contribute to the FUV and EUV flux at the
Western Wall (Fig.~\ref{fig:brg}), as do other O and B stars scattered throughout the region \citep{gagne11,wu18}.
The Wall has a particularly favorable geometry in that
it is a convex structure, which makes it easier to observe interface shapes than in
bowl-like concave interfaces like the Orion Bar \citep{goi19,goi16,pelligrini09}.
 
In what follows, we describe our image processing steps and 
combine the Br-$\gamma$, H$_2$, and K-cont AO images into a color composite in Sec.~2. 
The images reveal a new H$_2$ outflow and several new features along the interface, including long
ridges, sinusoidal-like waves, and fragments that we discuss in Sec.~3,
where we consider what the new observations tell us about instabilities, triggering, and
fragmentation processes.  A summary of the work is given in Sec.~4.

\section{Image Processing} \label{sec:data}

We acquired images of Carina's Western Wall with the Adaptive Optics Imager
(GSAOI) on Gemini South between Jan 15, 2018 and Jan 19, 2018 for a program with ten hours of queue time.
The images used three narrowband filters,
H$_2$ 1-0 S(1) (G1121; $\lambda_\circ$ = 2.122$\mu$m; $\Delta\lambda$ = 0.032$\mu$m),
Br-$\gamma$ (G1122; $\lambda_\circ$ = 2.166$\mu$m; $\Delta\lambda$ = 0.032$\mu$m), and
K-Continuum (G1112; $\lambda_\circ$ = 2.270$\mu$m; $\Delta\lambda$ = 0.034$\mu$m), and 
employed two pointings offset in declination from one another by 85\arcsec\ 
to cover the region of interest. The center of the composite image is
$\alpha$ = 10:43:30.6, $\delta$ = $-$59:35:20 (epoch 2000).  Within each pointing
we used multiple dithers of up to $\pm$ 8\arcsec\
to remove hot pixels and to fill in the $\sim$ 3\arcsec\ gaps
between the four Rockwell 2048 $\times$ 2048 arrays. Exposure times per dither
for the H$_2$, Br-$\gamma$, and Cont-K filters were 120, 120, and 100 seconds, respectively,
with 8 non-destructive reads for each exposure. The total exposure times in
the northern position were 1320, 1560, and 1400 seconds, respectively, for
H$_2$, Br-$\gamma$, and Cont-K, while the corresponding exposure times for the southern
position were 1920, 1680, and 1200 seconds. We took sets of dithered sky frames without AO-correction
and with the same exposure times as the object in
a relatively blank field located about 20 arcminutes away from the object. 
In all, the data set includes 56 sky frames, enabling us to correct for
changing atmospheric backgrounds in each filter as described below.

We combined dome and sky flats with the IRAF packages {\it gemcombine} and {\it gemarith}. For each
object frame we used the IRAF command {\it gemexpr} to flatfield the data, and
subtracted a sky composite acquired on the same night in that filter.
After using the `disco-stu' routines developed by the Gemini staff to correct for distortion,
we combined the dithers for each filter together with a median filter and
corrected for zero-point offsets in the background by
devising our own bad pixel masks for the data to eliminate residual
edge effects and bad columns. The two pointings then combine into a single
image by merging their data in the region of overlap. The resulting images have a plate
scale of 0.0195\arcsec\ per pixel, and cover a region 95\arcsec\ in RA by 173\arcsec\ in DEC.
The FWHM of the point-spread-function of the final images varies between 0.065\arcsec\ near
the center of each of the pointings, to 0.110\arcsec\ near the edges, where stellar
images become slightly distorted. These numbers compare well with the theoretical Rayleigh limit
of 0.069\arcsec\ for the Gemini 8-m telescope at 2.2$\mu$m. For comparison, JWST should 
achieve a resolution of 0.085\arcsec\ at this wavelength.

\begin{figure}
\plotone{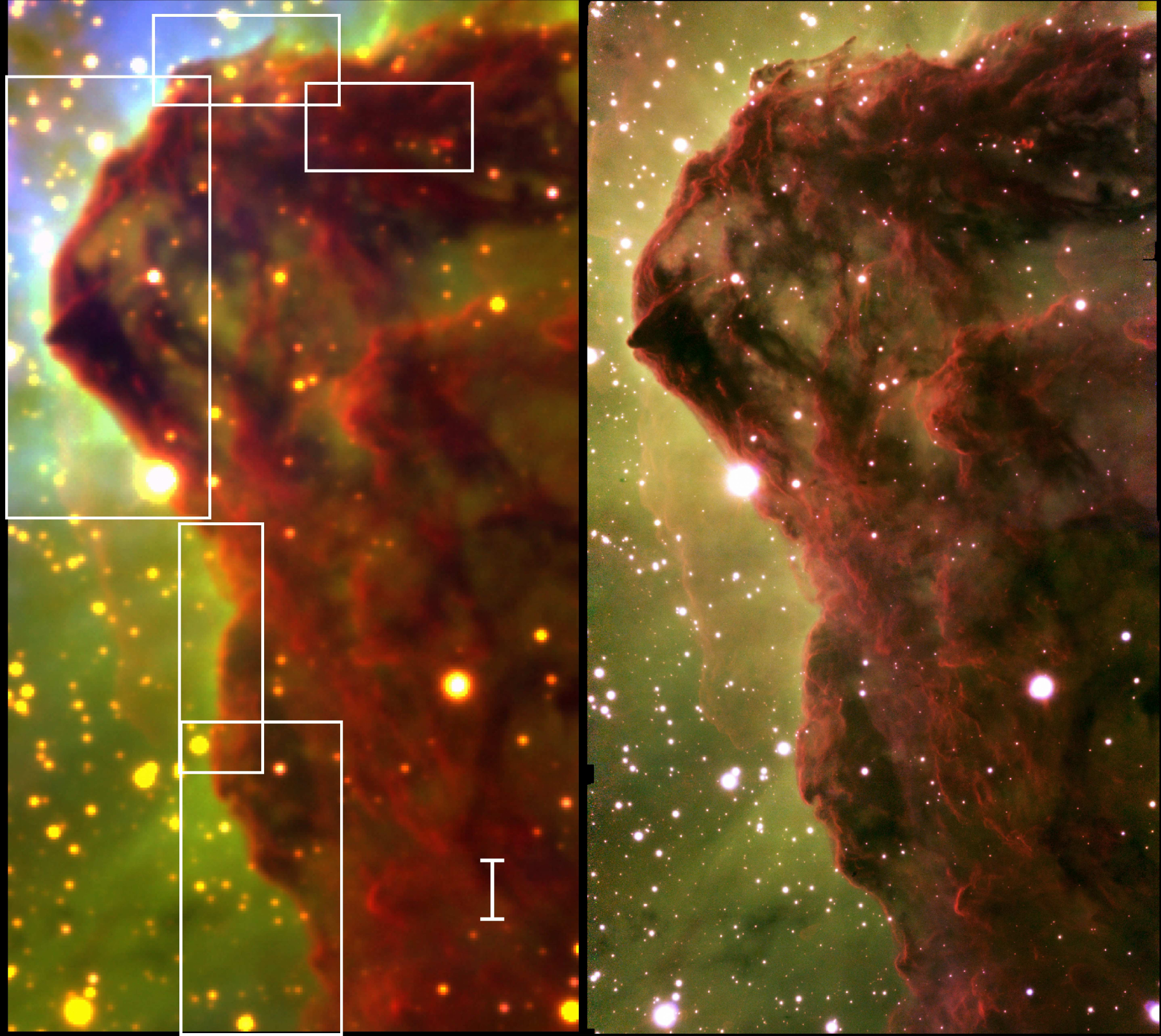}
\caption{Images of Carina's Western Wall. Left: A non-AO image of the region
from \citep{hartigan15} taken in 0.8\arcsec\ seeing, with H$_2$ in red, Br-$\gamma$ in green and
[O~III] $\lambda$5007 in blue. White boxes identify the regions expanded in Fig.~\ref{fig:zoom}. The scale bar is 0.1~pc
(8.97\arcsec).
Right: The new adaptive optics images acquired with GSAOI, where H$_2$ is in red, Br-$\gamma$ in green and K-cont
in blue. The FWHM of the stellar images ranges from 0.065\arcsec\ to 0.11\arcsec.}
\label{fig:overview}
\end{figure}

To improve the signal-to-noise we rebinned the AO images by 3$\times$3 pixels, so the images
we present here have square pixels of size 0.0585\arcsec. There are 80 stars in
the GAIA-DR2 catalog that have unsaturated profiles in our composite image, and the scatter in
their positions was rather large, $\pm$ 0.3\arcsec. Using the IRAF routines {\it geomap} and {\it geotran}
we redid the distortion corrections and were able to reduce the scatter in the 
GAIA-DR2 stellar coordinates by a factor of 15, to $\pm$ 0.02\arcsec\ in each of the three filters.
Fig.~\ref{fig:overview} compares our final AO composite with a ground-based non-AO image of the same
region.

\section{Discussion}

\subsection{Structure and Outflows Within The Western Wall}

\begin{figure}
\plotone{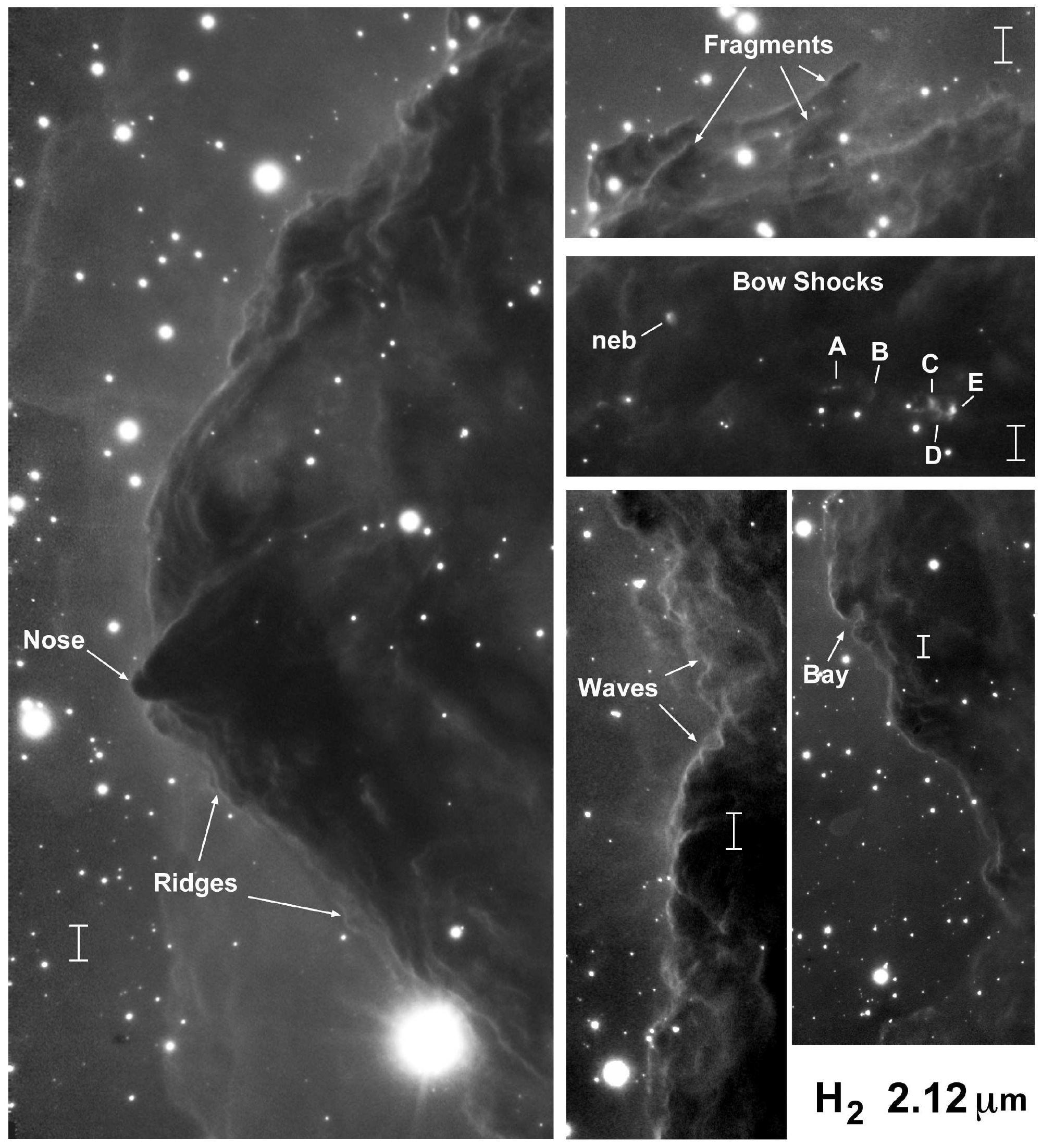}
\caption{Expanded views of sections of the H$_2$ adaptive optics image of Carina's Western Wall. Scale
bars are 5000~AU (2.17\arcsec), assuming a distance of 2.3~kpc.  Left: The apex of the Wall consists of
3 $-$ 4 long parallel ridges and a large triangular-shaped protuberance labeled `Nose'. 
Top-right: The northern edge of the Wall reveals several swept-back fragments along
the boundary of the object. Top-middle: An outflow is defined by a
series of H$_2$ bow shocks.  Bottom-right panels: The southern edge of the Wall
has a complex, and at times sinusoidal-like boundary and several overlapping arcs.}
\label{fig:zoom}
\end{figure}

The new AO images resolve previously unseen structure in the Western Wall
on scales between the seeing limit of the ground-based images (0.8\arcsec; 1840~AU) and
the diffraction limit of Gemini at 2.2$\mu$m (0.065\arcsec; 150~AU). The composite image in
Fig.~\ref{fig:overview} illustrates that this size range is particularly rich in detail
across the field of view, especially in H$_2$, which traces 
the photodissociation fronts.  In the H$_2$ image, the apex area of the Western Wall
appears as a series of long ridges that run along the interface, with projected
spacings of $\sim$ 2000~AU (Fig.~\ref{fig:zoom}).
A large triangular-shaped dark clump marked `Nose' 
protrudes from the wall near the apex. 
The northern boundary of the Wall appears more chaotic than the apex does, and 
is home to a series of swept-back fragments of width $\sim$ 1800~AU. To the south,
the Wall exhibits a remarkable set of wave-like interfaces, including
one that appears almost perfectly sinusoidal with a half-wavelength of 2500~AU.

From the color composite in Fig.~\ref{fig:overview} one might easily get the
impression that the Wall is transparent where H$_2$ emission is absent,
because Br-$\gamma$ is present between the filamentary H$_2$ emission.
However, in optical images the entire region where H$_2$ radiates is opaque,
so it is likely that the Br-$\gamma$ located within the Wall
comes from photoevaporation from its near-side in the direction of
the observer. If this interpretation is correct, this Br-$\gamma$ emission
lies superposed upon the H$_2$, and should be blueshifted relative to the
Br-$\gamma$ emission to the east of the Wall by $\sim$ 10 km$\,$s$^{-1}$,
the sound speed of the photoevaporative flow.

In addition to tracing fluorescence along irradiated interfaces, H$_2$ also emits
in shock waves along bipolar flows from stellar jets.
Our images reveal that an H$_2$ feature noted previously
as a possible shocked object \citep{tapia06}, resolves into
four nested bow shocks, labeled B, C, D, and E in Figure 3,
and a short linear feature, denoted A, located upstream (east)
of the bow shocks.  These objects differ from the irradiated interfaces in that they have no
associated Br-$\gamma$ or K-cont emission (pure red color in Fig~\ref{fig:overview}).
The region also stands out from the other H$_2$ emission in the non-AO ground-based images
\citep[][ left panel of Fig.~\ref{fig:overview}]{tapia06,hartigan15}, but with much less clarity.

The driving source of the jet is unknown. There is a narrow dark gap between knot
A and a faint nebula $\sim$ 0.6\arcsec\ to the east that could represent an
opaque disk, and another possibility for a driving source is the star labeled 'neb'
in Fig~\ref{fig:zoom} located 11.5\arcsec\ at PA 66.5 degrees from knot A. This star
has a conical-shaped reflection nebula aligned in the direction of the H$_2$ bow shocks,
but no nebulae connect the star with the H$_2$ shocks. The Vela-Carina
legacy survey with Spitzer did not detect any mid-infrared point
sources at the position of knot A, though the nebulous star is visible faintly
at 3.6 $\mu$m, 4.5 $\mu$m, and 5.8 $\mu$m. 
MHO~1630 is not obviously associated with a nearby filament of
faint optical emission lines \citep{tapia06}, or with any
sources visible in Spitzer or Herschel images \citep[e.g.][]{tapia15}.
Future ALMA mapping of the continuum and molecular line emission
in the vicinity of MHO~1630 would clarify whether or not protostars exist there.
Table~1 presents coordinates for several of the features shown in Fig.~\ref{fig:zoom}.

\begin{deluxetable}{lcc}
\tablewidth{0pt}
\tablenum{1}
\tablecolumns{5}
\tablecaption{Coordinates for Objects in Fig.~2}
\tablehead{}   
\startdata
\noalign{\smallskip}
Name&$\alpha$ (2000)&$\delta$ (2000)\\
neb&10:43:29.548&$-$59:34:10.695\\
MHO-1630 A&10:43:28.179&$-$59:34:15.154\\
MHO-1630 B&10:43:27.860&$-$59:34:15.570\\
MHO-1630 C&10:43:27.367&$-$59:34:16.132\\
MHO-1630 D&10:43:27.294&$-$59:34:16.805\\
MHO-1630 E&10:43:27.195&$-$59:34:16.578\\
Nose $^a$&10:43:35.813&$-$59:34:48.752\\
Wave $^{b, c}$&10:43:31.684&$-$59:34:33.942\\
Bay $^c$&10:43:31.772&$-$59:36:02.958\\
\noalign{\smallskip}
\enddata
\tablenotetext{a}{Apex location}
\tablenotetext{b}{Southern of the two waves marked in Fig.~\ref{fig:zoom}}
\tablenotetext{c}{Center of feature}
\end{deluxetable}

\subsection{Theoretical Implications}

The AO images in Figs.~\ref{fig:overview} and~\ref{fig:zoom}
allow us to observe the irradiated interfaces in the Western
Wall with unprecedented resolution, with implications for models of
triggered star formation.  In a simple one-dimensional model, we
expect the D-front associated with an irradiated interface to increase
the density in the cloud as the ionization front propagates, leading generally to 
conditions that are more favorable to gravitational collapse. In agreement
with this general picture, ALMA observations
along the southern portion of the Western Wall show that the densest portions of
the molecular cloud lie along the dissociation front \citep{hartigan20b}. 
In three dimensions, compression behind a D-front
becomes more complex.  On a timescale of $\sim$ 10$^5$ yrs, as a
front passes through clumps it will compress them,
and shadowing leaves behind a pillar with a dense knot at the apex which might then collapse
to form a new star \citep[e.g. Fig.~2 of][]{mackey10}.
Compression is highest when the front wraps around the clumps \citep{tremblin12},
and the external magnetic field can also influence the overall shape of the pillar \citep{henney09}.

If stars are indeed forming all along the Wall in response to the D-front, we would expect
pillars to trail behind the newly-formed stars away from the direction of 
radiation.  However, in contrast to the pillar-rich southern region of the Carina Nebula \citep{smith00}, 
the AO images do not show any indication that the Western Wall breaks into pillars
on size scales down to $\sim$ 150~AU.  Instead, we observe more of a smooth, slightly
wavy and sometimes ridged surface along much of the Wall down to the resolution of the images. 
The `Nose' in Fig.~\ref{fig:zoom} is the only feature where one can make a
case that radiation may be wrapping around a clump.  While this feature is striking and
may form one or more pillars at some point, it is currently a rather broad structure,
and it is unclear whether or not protostars or multiple cores exist there and, if
so, whether those clumps predated the passage of the front.
The available ALMA data on the Western Wall cover only its southern parts, and
the Nose has not been observed yet. Future high-resolution millimeter observations
might reveal the presence of protostars in this region.

The regularly-spaced ridges that run parallel to the photodissociation front
are a striking feature of the Western Wall.  Their geometry implies a preferred
direction for the ridges, characteristic of a dynamically significant large-scale
magnetic field.  Radiation-MHD simulations of H~II regions show that there should
be a large-scale magnetic field in the neutral gas that aligns along the
direction of the D-front \citep{arthur11}.
Comparing with the results of \citet{mackeylim11}, we see that their model R8
most closely resembles the ridges we see here. These simulations imply there should be
a large-scale dynamically significant magnetic field oriented northeast to southwest
in the Western Wall, parallel to the ridges. This prediction can be tested by
measuring the dust polarization along the Wall.

The flocculent nature of the northern edge of the Wall differs markedly
from the generally smoother, ridged morphology present near the apex. 
If the northern edge were strongly sculpted by the incident radiation
from Tr 14, then the dense structures here should form pillars that
point back towards the radiation sources to the east.  Instead, the northern
interface resembles one subject to Kelvin-Helmholtz (KH)
instabilities \citep{frank96,jones12}. If the
orientation of the global magnetic field is approximately constant, and
aligned as described above, then the field will be nearly perpendicular to the northern
edge and thus will not tend to stabilize the system against the KH instability.
The remaining ingredient for the KH instability is, of course, a shear flow. 
Radiation from nearby O stars in Tr~14 and Tr~16 is capable of causing just such a
shear flow along this edge and, indeed, the radiation itself may destabilize the
system further \citep[e.g.][]{shadmehri12}.

On a final note, it is particularly intriguing that the projected width of the
ridges referred to above, the scale of the density structures along the northern
edge of the Wall, and the wavelength of the sinusoidal structure referred to in
Sect 3.1 are all approximately 2000 AU in size. This is well within the
scales we would expect to be influenced by ambipolar diffusion in the undisturbed
molecular cloud, and therefore below the lengthscales expected to be produced by
turbulence \citep[e.g.][]{vanloo08,dos11,downes12,xu19}. Ambipolar
diffusion is usually significant only in weakly ionised regions and it is
therefore tempting to suppose that structures of this scale are created by flows
at the surface of the molecular cloud, where the ionisation fraction is high,
rather than being ``uncovered'' as photoevaporation of the cloud proceeds.

\section{Summary}

The H$_2$, Br-$\gamma$ and K-cont AO images Carina's Western Wall 
presented in this paper reveal several unexpected structures within
this classic irradiated molecular cloud that are difficult or impossible
to discern from non-AO ground-based images. The Wall has a convex geometry so
the interfaces appear as sharp tangent lines across the cloud, and are
ideal for resolving interface shapes that provide clues to MHD instabilities.
The apex of the Wall, which points in the general direction of the irradiating 
sources, consists of series of ridges that follow along the front. Wave-like
structures also occur along the front. The most chaotic morphologies occur along
the northern and southern edges of the Wall, where pillar-like fragments are
present. These fragments do not point towards the irradiating sources, but
are swept back as expected in a shear flow. 
The new images also resolve the H$_2$ object MHO~1630 into an outflow
that consists of a series of nested bow shocks without a clear driving source.

The observed ridging in the Wall argues for the presence of a large-scale
dynamically-important magnetic field, while the flocculent appearance along
the northern and southern boundaries is a characteristic of 
Kelvin-Helmholtz instabilities. We find no direct evidence that the ionization
front triggers new stars to form, in that there are no obvious cases where the
front wraps around a dense clump to form a protostar. Nonetheless, the
ionization front does encompass the entire Wall, and increases the
density along the surface where radiation is absorbed.

These observations make it clear that size scales between $\sim$ 200~AU and
2000~AU are rich with detail for an irradiated dark cloud, and in many ways
hold the key to understanding the complex dynamical processes that sculpt these
systems. The ridges, fragments, and waves we found all have sizes in this range,
and require subarcsecond resolution to resolve even for the nearest
regions of massive star formation. This capability now currently exists with
AO imaging, but should expand greatly once JWST becomes operational. 
The interplay between large-scale, global physics such as magnetic field
strengths and direction, rotation, and Jeans lengths, with the small scale three-dimensional
morphologies and complex motions induced by radiative feedback and turbulence
continues to make star formation a challenging and engaging field of study.

\acknowledgments
We thank Rodrigo Carrasco of the Gemini staff for his assistance with observation
scheduling and for routines we used for initial distortion corrections, and
D.~Froebrich for the official new MHO number.
Based on observations obtained for program GS-2018A-Q-123
at the international Gemini Observatory, a program of NSF’s NOIRLab,
which is managed by the Association of Universities for Research in Astronomy (AURA) under a
cooperative agreement with the National Science Foundation on behalf of the Gemini Observatory
partnership: the National Science Foundation (United States), National Research Council (Canada),
Agencia Nacional de Investigaci\'on y Desarrollo (Chile), Ministerio de Ciencia, Tecnologia e Innovacion
(Argentina), Minist\'erio da Ciência, Tecnologia, Inovacoes e Comunicacoes (Brazil), and Korea
Astronomy and Space Science Institute (Republic of Korea).

%

\vspace{5mm}
\facilities{Gemini (GSAOI)}

\end{document}